\documentclass[x11names,12pt]{article}
\usepackage[margin=.1in]{geometry}
\usepackage{jheppub} % for details on the use of the package, please see the JINST-author-manual
\usepackage[ragged]{footmisc}
\makeatletter
\renewcommand\section{\@startsection {section}{1}{\z@}%
                                 {-3.5ex \@plus -1ex \@minus -.2ex}%nn
                                   {2.3ex \@plus.2ex}%
                                   {\normalfont\large\bfseries}}
\renewcommand\subsection{\@startsection{subsection}{2}{\z@}%
                                   {-3.25ex\@plus -1ex \@minus -.2ex}%
                                     {1.5ex \@plus .2ex}%
                                     {\normalfont\bfseries}}
\renewcommand\subsubsection{\@startsection{subsubsection}{3}{\z@}%
                                   {-3.25ex\@plus -1ex \@minus -.2ex}%
                                     {1.5ex \@plus .2ex}%
                                     {\normalfont\itshape}}
\makeatother

\usepackage{titlesec}

\usepackage[utf8]{inputenc}
\usepackage{lmodern}
\usepackage[T1]{fontenc} % Use 8-bit encoding that has 256 glyphs
\usepackage{microtype} % Slightly tweak font spacing for aesthetics

\usepackage{xcolor}
\definecolor{dark-red}{rgb}{0.7,0,0}
\definecolor{dark-green}{rgb}{0.1,0.4,0}
\definecolor{NiceBlue}{rgb}{0.30196,0.55294,0.57647}
\usepackage{hyperref}
\hypersetup{
      colorlinks=true,
      linkcolor=dark-green,
      citecolor=dark-red,
      urlcolor=dark-green,
}
\newcommand{\bea}{\begin{eqnarray}}
\newcommand{\eea}{\end{eqnarray}}

\usepackage{mathtools}
\usepackage{enumitem}

\newcommand{\dd}{\mathrm{d}}

\begin{document}
%\begin{flushright}
%  \footnotesize \color{dark-green}{UUITP-xx/25}\\
%  \normalsize
%  \end{flushright}
\begin{titlepage}
  \thispagestyle{empty}

\vspace*{-0.8cm}
\begin{center}

{\bf {\LARGE \bfseries  Journey to the Center of a Black Hole}}

\begin{center}

\vspace{0.8cm}

{\bf  { Vyshnav Mohan} 
}

\vspace{0.2cm}
\hspace{.03em} The Oskar Klein Centre and Department of Physics, Stockholm University, \\
AlbaNova, 106 91 Stockholm, Sweden. \\
\vspace{0.5cm}
\hspace{.03em}Science Institute,
University of Iceland \\Dunhaga 3, 107 Reykjav\'{i}k, Iceland.

\vspace{0.2cm}

{\tt \href{mailto:vyshnav.vijay.mohan@gmail.com}{vyshnav.vijay.mohan@gmail.com}
}
\end{center}

\vspace{1.2cm}
{\bf Abstract}
\end{center}
\begin{quotation}
\noindent 
An infalling two-point function of a scalar field in a three-dimensional BTZ black hole, with one operator inserted in the exterior and the other inserted along a timelike geodesic, diverges when the infalling insertion point reaches the black hole singularity. In the boundary description, this divergence is tied to the exchange of double-twist operators rather than to the multi-stress-tensor sector that encodes curvature singularities in higher dimensions. We show that the inclusion of a tower of massive particles with Hagedorn growth gives rise to loop corrections that become important near the singularity, where they trigger a Hagedorn transition that renders the tree-level divergence untrustworthy. In the boundary theory, the corresponding tower of primary operators leads to a breakdown of the generalized free field description of the probe operator in the black hole state. In a large black hole, these corrections become important only near the singularity and remain exponentially suppressed at the horizon. The region where these corrections dominate is therefore parametrically separated from the horizon, as anticipated by the principle of black hole complementarity.

\end{quotation}
\vfill 

\end{titlepage}

\setcounter{page}{0}
\setcounter{tocdepth}{2}
\setcounter{footnote}{0}

\newpage
\pagestyle{plain}
\parskip 0.1in

\setcounter{page}{2}

\setcounter{tocdepth}{2}
%\hrule
{\hypersetup{linkcolor=black}
\tableofcontents
}
\afterTocSpace
\hrule
\afterTocRuleSpace

\section{Introduction}
What happens when you fall into a black hole? The principle of black hole complementarity proposes that an infalling observer sees a smooth horizon, even when quantum gravity effects are present \cite{Susskind:1993if}. This idea has been probed and prodded using various tools and techniques over the last few decades \cite{Almheiri:2012rt,Larjo:2012jt,Papadodimas:2012aq,Marolf:2013dba,Papadodimas:2013wnh,Papadodimas:2013jku,Lowe:2014vfa,Leutheusser:2021frk,Almheiri:2017fbd,Jafferis:2020ora,Gao:2021tzr,deBoer:2022zps,Franken:2026lgp,Coviello:2026tol,Coviello:2026hdc}.

In this paper, we study a sharp version of this question in a setting where the calculation is fully under control. We consider an `infalling' two-point function of a massive scalar field in a three-dimensional Bañados--Teitelboim--Zanelli (BTZ) black hole \cite{Banados:1992wn,Banados:1992gq}. One insertion point of the correlation function is fixed in the exterior of the black hole, while the other insertion point falls into the black hole along an infalling timelike geodesic. We find two types of divergences. The first occurs when the infalling insertion point reaches the singularity. To see the origin of this divergence, write the BTZ propagator as a sum over images of the AdS$_3$ propagator. As the singularity is approached, each term in the sum approaches the same constant. The sum therefore diverges (see also \cite{Hamilton:2007wj,Goto:2017olq}).\footnote{Finding signatures of black hole singularities both in the bulk and boundary correlation functions is a very active field of study. See \cite{Kraus:2002iv,Fidkowski:2003nf,Festuccia:2005pi,Hamilton:2007wj,Frenkel:2020ysx,Grinberg:2020fdj,Horowitz:2023ury,Dodelson:2023nnr,Ceplak:2024bja,Grozdanov:2026ktq} for examples.}

In the boundary theory, this divergence has a simple origin. Using the Hamilton--Kabat--Lifschytz--Lowe (HKLL) kernel, we can rewrite the bulk two-point function as an integral over a boundary correlation function \cite{Hamilton:2006fh,Hamilton:2006az}. At large $c$, the boundary correlator can be further written as the sum of the vacuum block and the exchange of the double-twist operators $[\mathcal{O}\mathcal{O}]_{n,j}$. If the spatial circle is decompactified, only the vacuum block survives and the bulk correlator reduces to an AdS$_3$ propagator, which is finite at $r=0$. The
divergence therefore requires the double-twist sector, whose role is to make the
correlator periodic on the spatial circle, as anticipated in~\cite{Ceplak:2024bja}. This is in contrast with higher dimensions, where the curvature singularity, which manifests through bouncing geodesics~\cite{Fidkowski:2003nf}, is encoded in the multi-stress-tensor sector of thermal correlators~\cite{Ceplak:2024bja}.

The second type of divergence arises when the exterior point becomes null separated from an image of the infalling point in the AdS$_3$ covering space. We refer to these as image singularities. Unlike the first type, they are not tied to the black hole singularity itself. Nevertheless, an infinite sequence of image singularities appears near the singularity and accumulates there.

The main goal of the paper is to understand which corrections become important as the infaller approaches the black hole singularity. We show that adding a tower of massive fields can introduce loop corrections to the propagator. In particular, we show that these loop corrections diverge as one approaches the singularity, completely dominating the semiclassical expression. The divergence of the loop amplitude is precisely analogous to that of a Hagedorn transition. In the standard Hagedorn story, the partition function on a thermal circle diverges when the size of the circle becomes small \cite{Sathiapalan:1986db,Kogan:1987jd,OBrien:1987kzw}. Here, the loop correction is controlled by the size of a loop attached to the infalling trajectory and its length goes to zero as one approaches the singularity. Having a tower of fields with an exponential density of states can then lead to a divergence of the total one-loop amplitude. What we have here is therefore a \emph{Lorentzian interior} Hagedorn transition. The boundary dual CFT does not have any such transition. It is only when the boundary theory is pulled into the bulk, and especially into the interior, that this Hagedorn transition appears.

This picture is closely related to earlier work on closed-string
tachyons. When a spatial circle shrinks to the string scale, strings
winding the circle can become tachyonic, and their condensation has been
proposed to replace spacelike
singularities~\cite{McGreevy:2005ci,Horowitz:2006mr}. Winding tachyons in
BTZ were studied in~\cite{Rangamani:2007fz}. In flat space, the Hagedorn
transition and the appearance of a tachyonic winding mode are two
descriptions of the same physics, related by modular
invariance~\cite{Atick:1988si}. Our analysis is complementary. We do not
assume a string completion, and work instead with an effective field
theory of a tower of particles with Hagedorn growth. This allows us to
compute where the transition happens relative to the horizon and to the
image singularities. Whether the
Hagedorn transition corresponds to an actual tachyon condensate depends on the
string completion,
and we return to this in Section~\ref{discussionsec}.

Including a tower of massive fields in the bulk corresponds to having a tower of primaries in the boundary conformal field theory (CFT). In the boundary theory, these corrections appear as new double-twist exchanges that eventually dominate over the large $c$ generalized free field description of the probe operator.

The image singularities can be shown to arise from the large frequency part of the bulk effective field theory (EFT). However, at finite $G_N$, or equivalently finite central charge $c$ in the boundary theory, the EFT approximation is not valid at such high energies/frequencies. Probe effects of the particle become important, and including these corrections resolves the image singularities.

Once we have identified these corrections, the natural question to ask is what happens at the horizon. For a large black hole, with a horizon radius much larger than the AdS length scale, we find that these corrections remain small when the infaller is near the horizon. The region where the semiclassical interior fails is parametrically separated from the horizon. This gives a concrete, controlled example in which corrections that must resolve the singularity do not propagate out to the horizon, in the spirit of black hole complementarity \cite{Susskind:1993if}.\footnote{For a small black hole, the location where these corrections become important depends on the Hagedorn growth rate. This leaves us with the theoretical possibility of infalling observers encountering `firewalls' in a small AdS black hole, as in \cite{Papadodimas:2012aq}.}

The rest of the paper is organised as follows. In Section \ref{divergencesec}, we study the infalling correlator and its divergences. In Section \ref{hagedornsec}, we show how a tower of new particles can lead to important loop effects near the singularity. In Section \ref{discussionsec}, we end with a discussion of our results and their implications for an infalling observer.

\section{Infalling Correlators and Trouble at the Singularity}
\label{divergencesec}
Consider a non-rotating BTZ black hole with the metric
\bea
\dd s^2 = -f(r)\dd t^2 + \frac{\dd r^2}{f(r)} + r^2 \dd\phi^2\,, \quad \text{where} \ f(r) = \frac{r^2-r_h^2}{L^2}\,.
\eea
We have denoted the AdS length scale by $L$. The inverse temperature of the black hole is given by 
\bea
\beta = \frac{2\pi L^2}{r_h}\,.
\eea
The BTZ black hole can be obtained from three-dimensional Anti-de Sitter space by taking the quotient of the spacetime under a discrete group of isometries generated by a boost. The AdS$_3$ spacetime can, in turn, be obtained by restricting four-dimensional flat spacetime $\mathbb{R}^{2,2}$, with metric
\bea
\dd s^2 = - \dd X_0^2 + \dd X_1^2 +\dd X_2^2-\dd X_3^2\,,
\eea
to the hyperboloid
\bea
-X_0^2 + X_1^2 +X_2^2- X_3^2 = -L^2\,.
\eea
The embedding coordinates are related to the BTZ coordinates through the relations \cite{Carlip:1995qv}
\bea
X_0=\frac{L r}{r_h}\cosh\left(\frac{r_h\phi}{L}\right),\qquad X_1=\frac{L r}{r_h}\sinh\left(\frac{r_h\phi}{L}\right)\,. \label{boosteq}
\eea

\begin{figure}
\centering
\includegraphics[width=0.35\linewidth]{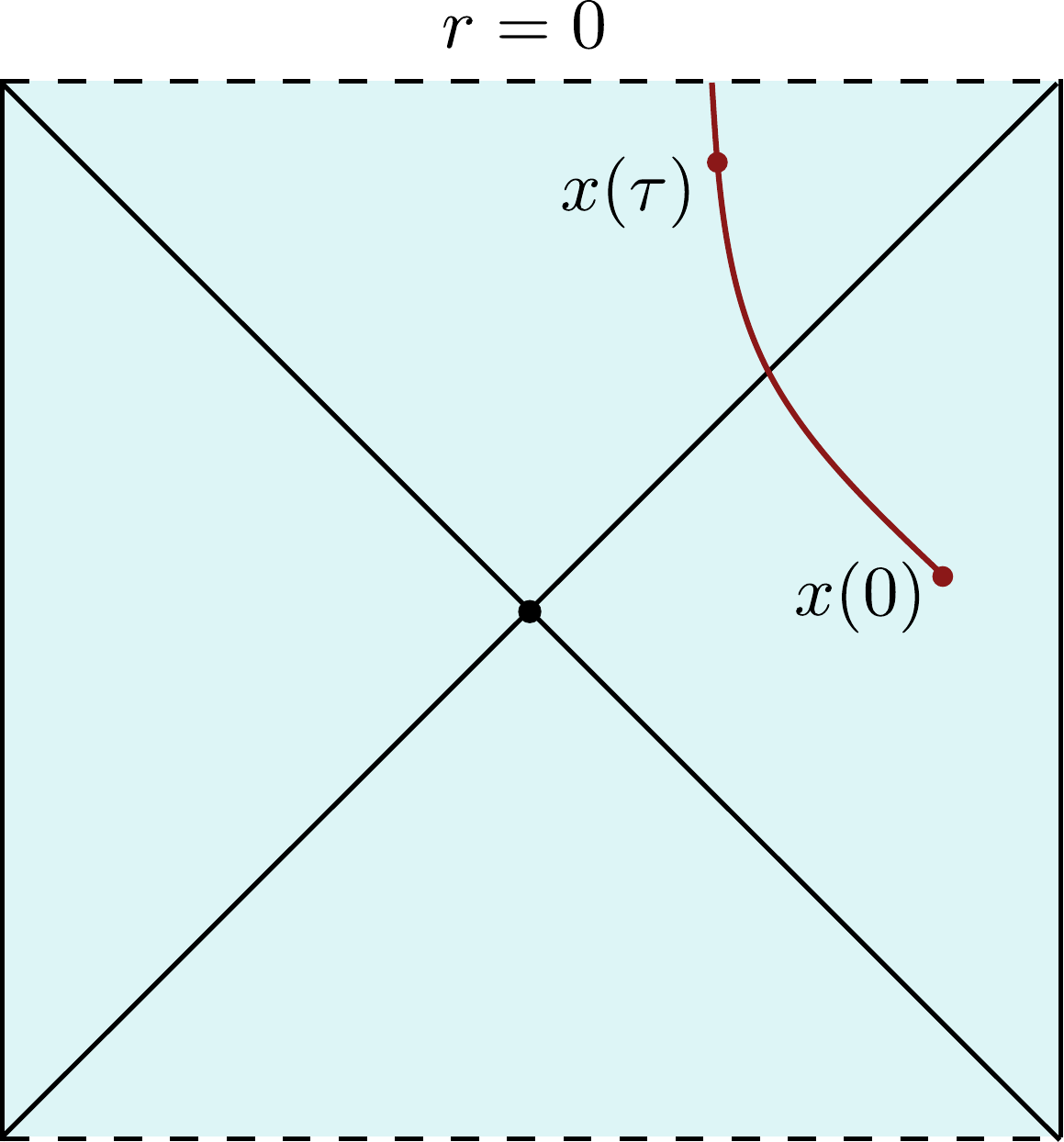}
\caption{\small An infalling observer in a BTZ black hole.}
\label{btzpenrosefig}
\end{figure}

\noindent When $r>r_h$, we have
\bea
X_2=L\sqrt{\tfrac{r^2}{r_h^2}-1}\cosh\left(\frac{r_ht}{L^2}\right),\ \qquad \ X_3=L\sqrt{\tfrac{r^2}{r_h^2}-1}\sinh\left(\frac{r_ht}{L^2}\right)\,.
\eea
On the other hand, when $r<r_h$, the coordinates are given by
\bea
X_2=L\sqrt{1-\tfrac{r^2}{r_h^2}}\sinh\left(\frac{r_ht}{L^2}\right),\ \qquad
X_3=L\sqrt{1-\tfrac{r^2}{r_h^2}}\cosh\left(\frac{r_ht}{L^2}\right)\,.
\eea

Now consider a timelike infalling radial geodesic starting from rest at a radius $r_0>r_h$ at some $t=t_0$. The equations of motion for the geodesic are given by
\bea
\begin{aligned}
p&=f(r)\dot{t}\,,\\
\dot{r}^2 &= p^2 - f(r)\,,
\end{aligned}
\eea
where $p$ is the conserved quantity along the geodesic and the derivatives are with respect to the proper time $\tau$. We have also chosen $\tau$ to vanish at the starting point. The radial equation can then be solved to give
\bea
r(\tau) = r_0\cos\left(\frac{\tau}{L}\right)\,.
\eea
The particle reaches the singularity after a finite proper time $\pi L/2$, irrespective of the value of $r_0$.

Now, let us assume that there is a scalar field $\Phi$ with mass $m$ living on the manifold. Consider an `infalling' correlator in the Hartle--Hawking vacuum,
\bea
G(\tau) := \left\langle \Phi(x(\tau))\Phi(x(0))\right\rangle_{\text{HH}}\,,\label{infallingcorreq}
\eea
where the points $x(\tau)$ are located along the infalling radial geodesic, as shown in Figure \ref{btzpenrosefig}. Using the method of images, we can rewrite this object in terms of the AdS$_3$ propagator as follows:
\bea
G(\tau)=\sum_{n \in \mathbb{Z}} G_{\mathrm{AdS}_3}\left(x(0), \gamma^n x(\tau)\right)\,.\label{propagatoreq}
\eea
For every point $x=(t,r,\phi)$ in AdS$_3$, the image $\gamma^n x$ is located at $(t,r,\phi+2\pi n)$.

The AdS$_3$ propagator is given by \cite{Hijano:2015qja,Grabovsky:2024jwf}
\bea
G_{\mathrm{AdS}_3}\left(x, x^{\prime}\right)=\frac{\left(\xi+\sqrt{\xi^2-1}\right)^{-\Delta+1}}{4\pi L \sqrt{\xi^2-1}}\,, \label{ads3propagator}
\eea
where $m^2L^2=\Delta(\Delta-2)$. It is convenient to express the function $\xi$ in the embedding coordinates as follows
\bea
\xi(X,Y) \equiv -\frac{X\cdot Y}{L^2},\qquad X\cdot Y = -X_0Y_0+X_1Y_1+X_2Y_2-X_3Y_3\,.
\eea
We also have the following relations (see for instance, \cite{Sarkar:2020yjs})
\bea
\begin{aligned}
\xi &= \cosh \left(\frac{\sigma}{L}\right)\,, &\text{(spacelike separated)}\,,\\
\xi &= 1\,, &\text{(null separated)}\,,\\
\xi &= \cos \left(\frac{\tau}{L}\right)\,, &\text{(timelike separated)}\,.
\end{aligned}
\eea
Here $\sigma$ is the geodesic distance between the two points, while $\tau$ is the proper time along the geodesic. In these cases, we can rewrite the AdS$_3$ propagator as follows
\bea
G_{\mathrm{AdS}_3}\left(x, x^{\prime}\right)=
\begin{dcases}
\frac{e^{-(\Delta-1)\sigma/L}}{4\pi \, L \, \sinh(\sigma/L)}\,, & \text{spacelike separation}\,.   \\
\frac{e^{-i(\Delta-1)\tau/L}}{4\pi i\,L\,\sin(\tau/L)}\,, & \text{timelike separation} \,. 
\end{dcases} \label{propagatortwocaseseq}
\eea

Now let us look at $\xi(x(0), \gamma^n x(\tau))$. When $n=0$, the two points $x(0)$ and $\gamma^n x(\tau)$ are timelike separated by construction. When $n \neq 0$, however, the two points need not be timelike separated, and this will lead to some interesting physics, as we will see in the next subsection. For simplicity, let us choose the value of the $\phi$ coordinate at $x(0)$ to be zero. Then we find
\bea
\xi_n \equiv \xi(x(0), \gamma^n x(\tau)) = \xi_0(\tau) + \frac{r_0\, r(\tau)}{r_h^2}\left[\cosh \left(\frac{2\pi n r_h}{L}\right)-1\right]\,, \label{nchordaldistance}
\eea
with $\xi_0(\tau) = \cos\left(\frac{\tau}{L}\right)$.

\subsection{Anatomy of a Fall}

As we approach the singularity, $\tau \to \pi L/2$. Then, from \eqref{nchordaldistance}, we can see that $\xi_n \to 0$ for all $n$. As a result, each term in the sum \eqref{propagatoreq} approaches the same constant,
\bea
\frac{1}{4\pi i L}\exp\left[-\frac{i\pi}{2}(\Delta-1)\right]\,.
\eea
The infalling correlator \eqref{propagatoreq} therefore diverges. An interesting observation is that this is not a UV divergence. Instead, it comes from the sum over images diverging. Understanding what kind of bulk and boundary effects can fix this divergence is the main goal of this paper.

Let us study this divergence in a little more detail. For some generic value $\tau \neq \pi L/2$, the sum over images can be split into two pieces: the range of $n$ for which $x(0)$ and $\gamma^n x(\tau)$ are timelike separated, and the range of $n$ for which the points are spacelike separated.\footnote{For certain special values of $\tau$, there can also be an integer $n$ for which $x(0)$ and $\gamma^n x(\tau)$ are null separated. We will return to this point later.} We can see from \eqref{propagatortwocaseseq} that the spacelike terms are exponentially suppressed relative to the timelike ones. In particular, for a fixed value of $\tau$, any term with
\bea
\xi_n>1  \quad \text{i.e.,} \quad |n| \gtrsim \frac{L}{2\pi r_h} \text{arcosh} \left(\frac{r(\tau) \left(r_0^2 -  r_h^2\right) + r_0 r_h^2}{r_0^2\, r(\tau)}\right)\,, \label{ncutoffeq}
\eea
is spacelike. Then these terms can be safely ignored from the sum \eqref{propagatoreq} (see also \cite{Hamilton:2007wj}). From the equation above, we see that more image terms become important as the infaller moves deeper into the interior. Near the singularity, all image terms contribute equally, causing the sum to diverge.

The propagator $G(\tau)$ also has another kind of divergence, which is not tied to the existence of the black hole singularity. For every value of $n\neq 0$, there exists a proper time $\tau$ at which $\xi_n=1$. This happens precisely when the following proper times are reached:
\bea
\tau_n = L\,\arccos \left[\frac{r_h^2}{r_0^2\cosh\!\left(\frac{2n\pi r_h}{L}\right) - r_0^2+r_h^2}\right]\,. 
\eea
This corresponds to the case where the points $x(0)$ and $\gamma^n x(\tau)$ are null separated. Geometrically, $\tau_n$ is the proper time at which a null geodesic emitted from $x(0)$ winds $n$ times around the $\phi$ circle before intersecting the infalling worldline at $x(\tau)$. We refer to these as \emph{image singularities} and they are bulk-to-bulk relatives of the bulk-cone singularities of boundary correlators \cite{Hubeny:2006yu,Dodelson:2020lal,Dodelson:2023nnr}, which arise from winding null geodesics between boundary points. Note that this divergence is of a completely different nature. It comes from the divergence of a single term in the sum \eqref{propagatoreq}, whereas the divergence at the singularity comes from the sum over all the terms.

For a large black hole, $r_h \gg L$, the location of these image singularities can be approximated as follows
\bea
\tau_n \approx \frac{\pi L}{2}-\frac{2L r_h^2 }{r_0^2} \exp\left[-\frac{2 \pi n r_h}{L}\right]\,, \qquad
r_n \approx \frac{2r_h^2}{r_0} \exp\left[-\frac{2 \pi n r_h}{L}\right]\,, \label{lightconeradius}
\eea
where $r_n$ is the radial coordinate of the $n$-th image singularity. We can see that these singularities lie deep inside the black hole interior and accumulate at the orbifold singularity.

Using these timescales, let us now describe the behaviour of the infalling correlator. At early proper times, from \eqref{ncutoffeq}, we can see that the correlator is dominated by the $n=0$ term. Nothing changes significantly even when the infaller crosses the horizon. However, when the infaller gets very close to the singularity, at $r(\tau_1)=r_1$, the correlator encounters the first image singularity. After this point, the first image term, $n=1$, becomes important. The infaller then hits the second image singularity at $r(\tau_2)=r_2$. After this proper time, the contribution from the second image term becomes important and can no longer be neglected. This continues as the infaller approaches the singularity. More and more image terms become important, until eventually the infaller reaches the singularity, where \emph{all} image terms contribute equally, and the correlator diverges.

\section{A Bulk Hagedorn Transition}
\label{hagedornsec}

In a theory of quantum gravity, we do not expect the divergences associated with the singularity to be present. In this section, we identify the corrections beyond the probe EFT that lead to the resolution of such divergences. Before turning to the problem at hand, it is instructive to consider another example. Suppose that the points in the correlator \eqref{infallingcorreq} are not located along an infalling geodesic, but instead lie along a constant $r$ curve in the right exterior of the black hole. The accelerated `observer' then stays near the horizon for an infinite proper time. As a result, the observer has enough energy resolution to see the discreteness of the black hole states. At times of order the exponential of the Bekenstein--Hawking entropy of the black hole, we therefore expect the semiclassical description of the correlator to break down. This breakdown is directly related to the resolution of Maldacena's information loss problem in the boundary theory \cite{Maldacena:2001kr}. The important takeaway for us is that these corrections are not important for our infalling observer, because they reach the singularity in a finite proper time. Consequently, an infalling observer can only probe energies much larger than the energy level spacing of the black hole microstates.

\begin{figure}
\centering
\includegraphics[width=0.3\linewidth]{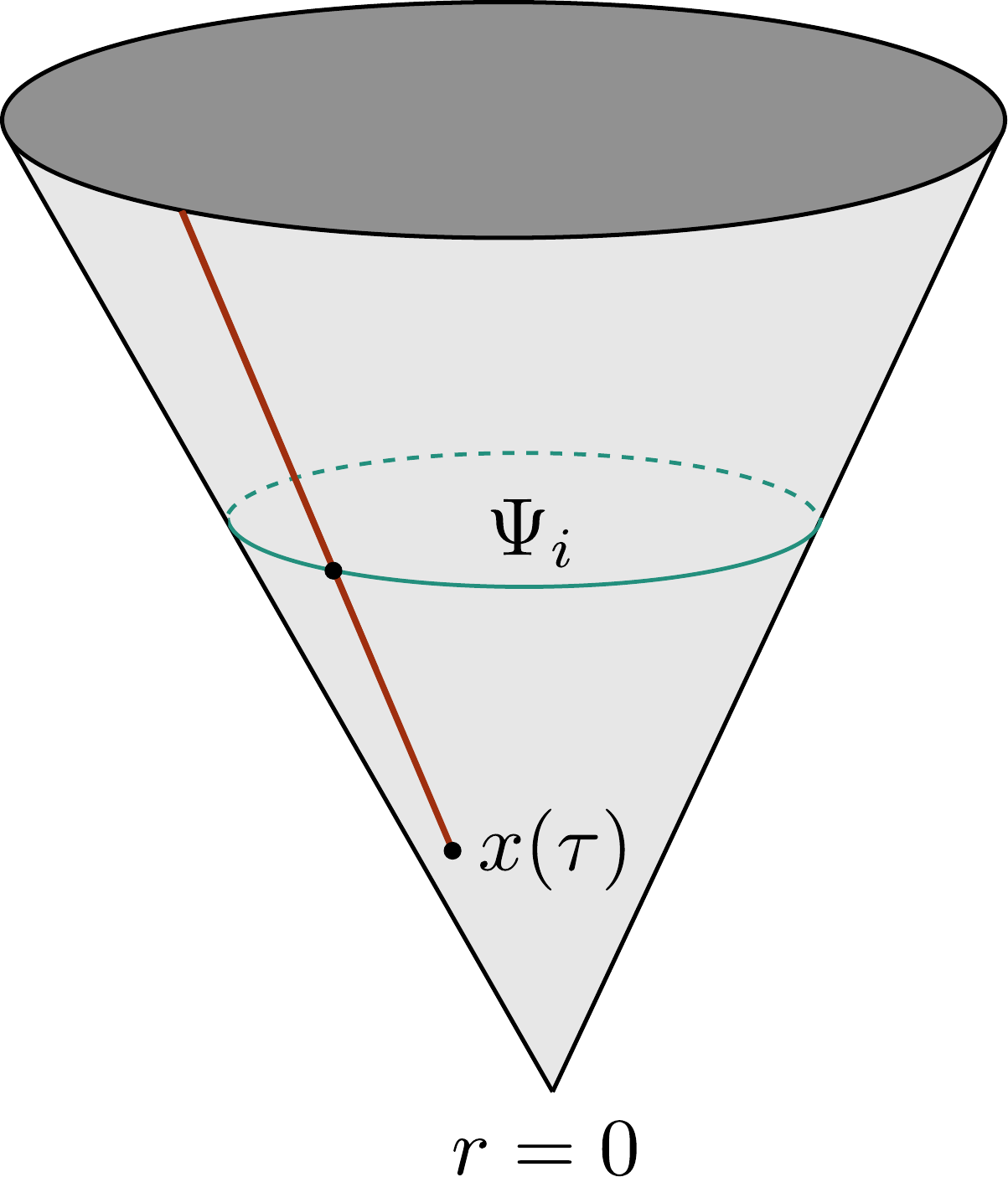}
\caption{\small A field $\Psi_i$ from the tower produces a one-loop correction to the $\Phi$ propagator in which a loop winding the $\phi$ circle is attached along the trajectory. The dominant contribution comes from a loop attached at $x(\tau)$, whose geodesic length $\ell(\tau)$ shrinks to zero as $\tau \to \pi L/2$.}
\label{windingfig}
\end{figure}

It is therefore interesting to ask which corrections are important to an infalling observer. Let us begin by assuming that, in addition to the particle associated with the scalar field $\Phi$, there exists a tower of particles with masses $m_i$. We will model these particles as an EFT of new scalar fields $\Psi_i$, coupled to $\Phi$ through
\bea
\mathcal{L}_{\rm int} = -\,g \sum_i \Phi^2 \Psi_i^2\,.
\eea

These interactions can result in loop corrections to the propagator of the field $\Phi$. In particular, our infalling correlator receives a correction in which, at some point along the trajectory, a loop of one of the fields $\Psi_i$ can be attached (see Figure \ref{windingfig} for a schematic representation of such a correction). The correction from such a loop is given by
\bea
g\int \dd y \ G_{\text{bulk}}(x(0),y)\langle\Psi^2_i(y)\rangle_{\text{ren}} G_{\text{bulk}}(y,x(\tau))\,. \label{bulkloopdiagram}
\eea
The loop propagator of the field $\Psi_i$ can once again be computed using the method of images:
\bea
\langle\Psi^2_i(y)\rangle_{\text{ren}} = \sum_{n \neq 0 } G_{\mathrm{AdS}_3, \Psi_i}\left(y, \gamma^n y\right)\,.
\eea
Note the $n=0$ term is not included in the sum because it diverges and is absorbed by the renormalization of the mass of the field $\Phi$. For $m_i L \gg 1$, the $n=\pm1$ terms dominate, and we have
\bea
\langle\Psi^2_i(y)\rangle_{\rm ren} \sim e^{-m_i \sigma_1(y)}\,, \qquad \text{where} \ \sigma_n(y) \equiv \sigma(y, \gamma^n y) \,.
\eea
Using the embedding coordinates of the previous section, we can show that for points $y$ both inside and outside the horizon,
\bea
\sinh \left(\frac{\sigma_n(y)}{2 L}\right) = \frac{r_y}{r_h} \sinh \left(\frac{n\pi r_h}{L}\right)\,, \label{selfimagedistance}
\eea
where $r_y$ is the value of the $r$ coordinate at $y$. At the horizon, this gives $\sigma_1(y) = 2\pi r_h$, as expected. In the interior $\sigma_1$ decreases monotonically with $r$ and vanishes at $r=0$.
 
We can now identify the dominant contribution to \eqref{bulkloopdiagram}. The intermediate point $y$ must lie in the past lightcone of the future insertion point $x(\tau)$ (see Appendix \ref{ininapp} for details). The further the loop is from $x(\tau)$, the larger its length is. The dominant correction therefore comes from a loop attached to the future insertion point $x(\tau)$, with length
\bea
\ell(\tau) \equiv \sigma_1\big(x(\tau)\big) = 2L\, \operatorname{arcsinh}\!\left[\frac{r(\tau)}{r_h} \sinh \left(\frac{\pi r_h}{L}\right)\right] \,. \label{loopdistanceeq}
\eea
Note that if the location of the intermediate point $y$ were unrestricted, we could envision a situation in which the leading contribution comes from a configuration where $y$ lies at the future or past singularity. See Appendix \ref{ininapp} for more details on why these configurations are not allowed. Therefore, the correction from the field $\Psi_i$ is
\bea
\delta G_i(\tau) \sim g\, G_{\text{tree}}(\tau)\, e^{-m_i \ell(\tau)}\,,
\eea
where $G_{\text{tree}}$ is the tree-level propagator \eqref{propagatoreq}. The total one-loop contribution coming from the entire tower of fields is then given by
\bea
\delta G(\tau) = \sum_i \delta G_i(\tau)\,.
\eea
We can replace the sum over particles with an integral over the mass by introducing a density $\rho(m)$:
\bea
\frac{\delta G(\tau)}{G_{\text{tree}}(\tau)} \sim g \int \dd m \ \rho(m)\, e^{-m \ell(\tau)}\,. \label{totalloopamplitude}
\eea
 
\noindent Now let us \emph{choose} the density of states to have Hagedorn growth,
\bea
\rho(m) = e^{\beta_H m}\,. \label{hagedrongrowtheq}
\eea
One might think that exponential growth of the density of states is a
restrictive, or even exotic, assumption. However, if the tower comes from
a string theory in the bulk, as we discuss in Section~\ref{boundarysec},
this is a very natural choice.

Now let us make a very interesting observation. As $\tau \to \pi L/2$, we have $\ell(\tau) \to 0$. Therefore, there exists a critical value of $\ell(\tau)$,
\bea
\ell_H = \beta_H\,, \label{hagedorntemp}
\eea
below which the integral \eqref{totalloopamplitude} diverges.\footnote{At finite $c$, the tower is cut off at the black hole threshold and this divergence becomes a sharp crossover. See the end of
Section~\ref{boundarysec} for more details.} This is precisely a Hagedorn transition, but occurring in the interior of a black hole.

We can use \eqref{loopdistanceeq} to show that an infalling observer hits the Hagedorn \emph{wall} at a proper time
\bea
\tau_H =L \arccos \left[\frac{r_h \sinh \left(\frac{\beta_H }{2 L}\right)}{r_0\sinh\left(\frac{\pi r_h}{L}\right)}\right] \,,
\eea
at the Hagedorn radius
\bea
r_H = r_h \sinh \left(\frac{\beta_H}{2 L}\right) \text{csch}\left(\frac{\pi r_h}{L}\right)\,.  \label{fullhagedornradius}
\eea
For a large black hole, we can approximate these quantities as follows
\bea
\tau_H \approx \frac{\pi L}{2} -\left(\frac{2 r_h L\sinh \left(\frac{\beta_H }{2 L}\right)}{r_0}\right)\exp\left[-\frac{\pi r_h}{L}\right]\,, \quad r_H \approx 2 r_h \sinh \left(\frac{\beta_H }{2 L}\right)\exp\left[-\frac{\pi r_h}{L}\right]\,.\label{hagedornradiusexpression}
\eea
Therefore, the tree-level propagator cannot be trusted beyond $\tau_H$.
In particular, it cannot be trusted near the singularity, precisely where
it diverges.

\subsection{Boundary interpretation}
\label{boundarysec}
 
Now that we have a bulk EFT description of the corrections, let us ask what they correspond to in the boundary theory. We begin by rewriting the bulk two-point function in terms of boundary correlators using the HKLL construction \cite{Hamilton:2005ju,Hamilton:2006az}. In the right exterior wedge, we can expand the field in Schwarzschild modes as follows
\bea
\Phi_R(x)=\sum_{q\in\mathbb{Z}}\int_0^\infty \dd\omega \left(f_{\omega q}(r)\, e^{-i\omega t + i q\phi}\, a_{\omega q}^{(R)}+ \text{h.c.}\right)\,, \qquad x=(t,r,\phi)\,,
\eea
where $f_{\omega q}(r)$ solves the radial Klein--Gordon equation with the normalisable boundary condition
\bea
f_{\omega q}(r)\to r^{-\Delta} \  \text{as}  \ r\to\infty\,.
\eea
The boundary operator dual to $\Phi$ is then given by the standard extrapolate dictionary
\bea
\mathcal{O}_R(t,\phi) = \lim_{r\to\infty} r^{\Delta}\,\Phi_R(x) = \sum_{q}\int_0^\infty \dd\omega \left( e^{-i\omega t + i q\phi}\, a_{\omega q}^{(R)}+ \text{h.c.}\right)\,.
\eea
The creation and annihilation operators can be obtained by inverting the Fourier transform,
\bea
a^{(R)}_{\omega q} = \frac{1}{(2\pi)^2}\int \dd \tilde t\, \dd\tilde\phi\; e^{i\omega \tilde t - i q\tilde \phi}\, \mathcal{O}_R(\tilde t,\tilde\phi)\,, \qquad \omega>0\,.
\eea
Substituting this expression back gives the exterior HKLL representation
\bea
\Phi_R(x)=\int \dd^2 \tilde{x} \ K^{(R)}_{\text{ext}}\left(x; \tilde{x}\right) \mathcal{O}_R(\tilde{x})\,, 
\eea
where 
\bea
K^{(R)}_{\text{ext}}\left(x ; \tilde{x} \right)=\frac{1}{(2\pi)^2}\sum_{q}\int_0^\infty\dd\omega\left[ f_{\omega q}(r)\, e^{-i \omega (t-\tilde t)+i q(\phi-\tilde\phi)} + \text{c.c.}\right]\,.\label{HKLLeq}
\eea

For a point $x_2$ in the future interior, the field depends on the Cauchy data on both exterior regions. Continuing the right and left exterior modes across the future horizon \cite{Hamilton:2006az,Papadodimas:2012aq} one has
\bea
\Phi_{\rm int}(x_2) = \sum_{q}\int_0^\infty\dd\omega\left(g^{(R)}_{\omega q}(x_2)\, a^{(R)}_{\omega q}+g^{(L)}_{\omega q}(x_2)\, a^{(L)}_{\omega q} + \text{h.c.}\right)\,,
\eea
where $g^{(R,L)}_{\omega q}$ are the interior continuations of the exterior mode functions, and the same Fourier inversion on each boundary gives
\bea
\Phi_{\rm int}\left(x_2\right)=\int \dd^2 \tilde{x} \left[K_{\mathrm{int}}^{(R)}\left(x_2 ; \tilde{x} \right) \mathcal{O}_R(\tilde x)+K_{\mathrm{int}}^{(L)}\left(x_2 ; \tilde{x} \right) \mathcal{O}_L(\tilde x)\right]\,,
\eea
with $K^{(R,L)}_{\rm int}$ given by \eqref{HKLLeq} with $f_{\omega q}$ replaced by $g^{(R,L)}_{\omega q}$. The interior HKLL kernels are related to the exterior ones by suitable analytic continuations. The infalling correlator is therefore given by the expression
\bea
\begin{aligned}
G(\tau) = \left\langle\Phi\left(x(\tau)\right) \Phi\left(x(0)\right)\right\rangle &=\int \dd^2\tilde{x}\, \dd^2 \tilde{y} \ \Big[K_{\mathrm{int}}^{(R)}\left(x(\tau) ; \tilde{x} \right) \left\langle\mathcal{O}_R(\tilde{x} ) \mathcal{O}_R(\tilde{y} )\right\rangle\\
&\qquad\qquad +K_{\mathrm{int}}^{(L)}\left(x(\tau) ; \tilde{x} \right)\left\langle\mathcal{O}_L(\tilde{x} ) \mathcal{O}_R(\tilde{y} )\right\rangle\Big]\, K^{(R)}_{\text{ext}}\left(x(0) ; \tilde{y} \right)\,.
\end{aligned}
\eea
The two-sided correlator in the thermofield double (TFD) state is a single-sided thermal correlator at a shifted complex time,
\bea
\left\langle\mathcal{O}_L(t, \phi) \mathcal{O}_R\left(t^{\prime}, \phi^{\prime}\right)\right\rangle_{\mathrm{TFD}}=\left\langle\mathcal{O}_R(-t + i \beta / 2, \phi) \mathcal{O}_R\left(t^{\prime}, \phi^{\prime}\right)\right\rangle_\beta\,.
\eea
This gives us
\bea
G(\tau)=\int \dd^2\tilde{x}\, \dd^2 \tilde{y} \ K\left(x(\tau), x(0) ; \tilde{x} , \tilde{y} \right)\langle\mathcal{O}(\tilde{x} ) \mathcal{O}(\tilde{y} )\rangle_\beta\,, \label{bulkfromboundary}
\eea
with
\bea
K\left(x(\tau), x(0) ; \tilde{x} , \tilde{y} \right)=\left[K_{\mathrm{int}}^{(R)}\left(x(\tau) ; \tilde{x}\right)+K_{\mathrm{int}}^{(L)}\left(x(\tau) ; -t_{\tilde{x}}+\tfrac{i \beta}{2}, \phi_{\tilde{x}}\right)\right] K^{(R)}_{\text{ext}}\left(x(0) ; \tilde{y}\right)\,,
\eea
where $(t_{\tilde{x}}, \phi_{\tilde{x}})={\tilde{x}}$.

Now, let us rewrite the thermal trace as a sum over energy eigenstates. Above the Hawking--Page transition, this sum is dominated by states $|H\rangle$ with $E_H\sim c$. Therefore, we have
\bea
G(\tau)\approx\frac{1}{Z}\sum_{H}e^{-\beta E_H}\int \dd^2\tilde{x}\, \dd^2 \tilde{y} \ K\left(x(\tau), x(0) ; \tilde{x} , \tilde{y} \right)\left\langle H |\mathcal{O}(\tilde{x}) \mathcal{O}(\tilde{y})|H\right\rangle\,.
\eea
The bulk correlator is thus an integral over heavy--heavy--light--light (HHLL) correlators, which admit the following conformal block expansion in the $\mathcal{O}\times\mathcal{O}$ channel:
\bea
\begin{aligned}
\left\langle O_H(\infty) O_L(1) O_L(z) O_H(0)\right\rangle & =\sum_{\text {primaries } p} C_{L L p} C_{H H p}\,\mathcal{F}\left(h_L, h_H, h_p, c, z\right)\bar{\mathcal{F}}\left(\bar h_L, \bar h_H, \bar h_p, c, \bar z\right)\,, \\
& \simeq \int \dd{\Delta_p} \ \rho\left(\Delta_p\right) \overline{C_{L Lp} C_{H Hp}} \ \mathcal{F}\bar{\mathcal{F}}\,.\label{cfbexpansion}
\end{aligned}
\eea
In the second line, the OPE coefficients are averaged over Virasoro primaries of dimension $\Delta_p$ \cite{Kraus:2016nwo,Michel:2019vnk}, and $\rho(\Delta_p)$ is their density.
 
Now let us investigate which conformal blocks contribute to the correlation function in
the large $c$ limit. The primary suspect is the vacuum block, which corresponds to the
exchange of the identity primary and its Virasoro descendants. This exchange `builds'
the BTZ background on which the light operator propagates as a free particle
\cite{Fitzpatrick:2015zha}.

Periodicity of the spatial circle also requires the exchange of the double-twist
operators $[\mathcal{O}\mathcal{O}]_{n,j}$~\cite{Grabovsky:2024jwf,Ceplak:2024bja}. The
vacuum block together with the $[\mathcal{O}\mathcal{O}]_{n,j}$ exchanges constitutes
the large $c$ generalized free field description of $\mathcal{O}$ in the heavy state
(see also \cite{El-Showk:2011yvt,Heemskerk:2009pn,Iliesiu:2018fao}). In the bulk, this
description corresponds to the free field $\Phi$ propagating on the BTZ background.

Now let us make an interesting observation. Without the $[\mathcal{O}\mathcal{O}]_{n,j}$
exchanges, the boundary correlator is the vacuum block alone, which is the thermal
correlator on the decompactified (planar) geometry. The bulk correlator is then just the
$n=0$ term $G_{\mathrm{AdS}_3}(x(0),x(\tau))$, which is finite at $r=0$. The divergence
at $r=0$ therefore comes from the $[\mathcal{O}\mathcal{O}]_{n,j}$ exchanges.

This is the $d=2$ counterpart of the results of~\cite{Ceplak:2024bja}. For $d\geq 3$,
signatures of the curvature singularity, through the bouncing geodesics, reside in the
multi-stress-tensor sector of thermal correlators. In $d=2$, the vacuum block, which
contains the multi-stress-tensor exchanges, by itself describes a geometry without a
singularity, and the divergence can be attributed to the
$[\mathcal{O}\mathcal{O}]_{n,j}$ sector.

Let us now examine what the loop corrections discussed above correspond to in the boundary theory. The bulk tower translates into a tower of primaries $\mathcal{O}_i$ with $\Delta_i \simeq m_i L$. The one-loop corrections then correspond to the exchange of the double-twist operators $[\mathcal{O}_i\mathcal{O}_i]_{n,j}$ with structure constants $C_{LL[\mathcal{O}_i\mathcal{O}_i]}\sim g$. These exchanges would then show up as new corrections to the generalized free field description.
 
We assumed Hagedorn growth \eqref{hagedrongrowtheq} for these light primaries. This is compatible with the Hartman--Keller--Stoica (HKS) sparseness condition \cite{Hartman:2014oaa} 
\bea
\rho(\Delta)\lesssim e^{2\pi\Delta}\,, \qquad \text{for} \ \ \Delta\leq c/12 \,,
\eea
provided
\bea
\beta_H \lesssim 2\pi L\,. \label{HKSbound}
\eea
The HKS bound also tells us where Hagedorn growth must stop. Above the
threshold $\Delta\simeq c/12$, the density of states is governed by a slower
Cardy formula, 
\bea
\rho(\Delta)\sim\exp\left(2\pi\sqrt{\frac{c\Delta}{3}}\right)\,.
\eea
In the bulk, states with
$m\gtrsim c/(12L)$ correspond to BTZ black holes rather than particles and are not
described by the probe EFT. Cutting off the tower at
$m_{\max}\simeq c/(12L)$, we find, up to power-law prefactors in $m$ and
$L$,
\bea
\frac{\delta G(\tau)}{G_{\rm tree}(\tau)}
\sim \frac{g}{\beta_H-\ell(\tau)}\,
\exp\!\left[\frac{c}{12L}\big(\beta_H-\ell(\tau)\big)\right]\,.
\eea
At finite $c$, the Hagedorn transition is therefore not a true divergence but a crossover. The one-loop perturbation theory in $g$ breaks down once $|\delta G/G_{\rm tree}|\sim 1$, that is,
\bea
\beta_H-\ell(\tau)\gtrsim -\frac{12L}{c}\log\left(\frac{cg}{12L}\right)\,.
\eea

An important point to note here is that the Hagedorn transition we saw in the interior is absent in the boundary theory. The boundary correlator itself is, of course, finite and the `transition' is a statement about the bulk EFT reorganisation of the sum over primaries after integration against the kernel $K$ in \eqref{bulkfromboundary}. For $r(\tau)<r_H$, the double-twist contributions of the tower are no longer suppressed relative to the vacuum block and the double-twist $[\mathcal{O}\mathcal{O}]_{n,j}$ exchanges, and in this sense, the generalized free field description breaks down.

Hagedorn growth strongly suggests that the tower comes from a string theory in the bulk (see for example \cite{Belin:2014fna}). As a simple example, let us model the tower using closed bosonic string theory. The first massive state has
\bea
\Delta_{\text {gap}} \simeq m_1 L=\frac{2 L}{\sqrt{\alpha^{\prime}}}\,,
\eea
and the level-$N$ degeneracy $d_N d_{\tilde{N}}$, with $d_N \sim \exp\big(2\pi\sqrt{c_{\text{eff}}N/6}\big)$, $c_{\text{eff}}=24$, and $N=\alpha' m^2/4$, gives
\bea
\beta_H = 4\pi \sqrt{\alpha^{\prime}} = \frac{8\pi L}{\Delta_{\rm gap}}\,, \label{spectralgapbetarelation}
\eea
so that \eqref{HKSbound} becomes $\Delta_{\rm gap}\gtrsim 4$. A single boundary number, the gap to the massive tower, therefore fixes the location \eqref{hagedornradiusexpression} of the Hagedorn wall in the bulk. It would be interesting to explore whether this condition can be incorporated as an additional bootstrap constraint on holographic CFTs dual to BTZ black holes, extending the set of conditions considered in \cite{Heemskerk:2009pn,El-Showk:2011yvt,Hartman:2014oaa}.

\subsection{Resolving the image singularities}
\label{imagesiingularitysec}
\begin{figure}
\centering
\includegraphics[width=1\linewidth]{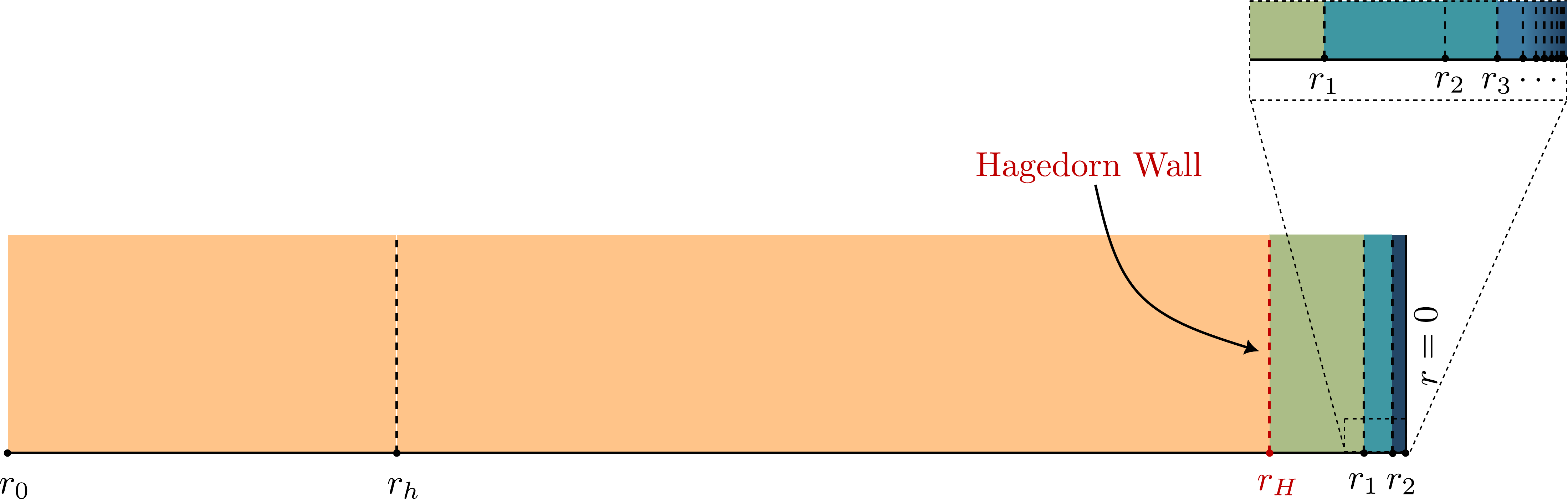}
\caption{\small The location of the Hagedorn wall and the image singularities in a large BTZ black hole and sufficiently large $\beta_H$. The Hagedorn transition occurs before any of the image singularities show up.}
\label{infalltimefig}
\end{figure}
Let us now return to the image singularities of Section \ref{divergencesec}.  The coincident-point lightcone singularity of the $n=0$ term is fixed by the operator product expansion and is present in the exact theory at any $c$. The image singularities are different. These singularities show up only when we integrate the $c=\infty$ generalized free field correlator over the bulk HKLL kernel $K$, and nothing in the exact finite-$c$ theory requires them. In this respect, they are the Lorentzian analogue of the forbidden singularities of the semiclassical heavy--light Virasoro block \cite{Fitzpatrick:2016ive}. They are both artifacts of the $c \to \infty$ limit and are absent in a finite $c$ theory.
 
One can estimate how these image singularities are resolved in the finite $c$ theory. Let us expand the AdS$_3$ propagator in global modes,
\bea
G\left(X_1, X_2\right)=\sum_{n \geq 0, j \in \mathbb{Z}} e^{-i \omega_{n j}\Delta t+i j \Delta \phi} R_{n j}\left(\rho_1\right) R_{n j}\left(\rho_2\right)\,, \quad \omega_{nj} L = \Delta + |j|+2n\,, \label{porpagatormodeexpansion}
\eea
We can split this sum at some frequency $\omega_{nj} = \Lambda$:
\bea
G = G_{<\Lambda}+G_{>\Lambda}\,.
\eea 
The sum $G_{<\Lambda}$ is a bounded and truncated Fourier series. Therefore, the $1/\sqrt{\xi_n^2-1}$ singularity of an image term is produced entirely by modes present in $G_{>\Lambda}$.

At finite $c$, however, the generalized free field description of $\mathcal{O}$ that underlies the image sum fails for energies above the black hole threshold, $\omega L\gtrsim c$ \cite{Hartman:2014oaa}. In other words, the probe approximation for $\Phi$ breaks down at large frequencies. Therefore, the modes responsible for the singularity are not present in the finite $c$ theory. Truncating the sum at $\Lambda\sim c/L$ smears the singularity over a width $|\xi_n-1|\sim 1/c$. Interestingly, forbidden singularities in the boundary Euclidean theory also get resolved by a similar mechanism \cite{Fitzpatrick:2016ive,Faulkner:2017hll}. Note also that this truncation does nothing to the divergence at $r=0$, which comes from the low-frequency, $\xi_n\to0$ regime of every image term.

 Remarkably, for a large black hole ($r_h \gg L$) and sufficiently large $\beta_H$, all the image singularities lie behind the Hagedorn wall. Comparing \eqref{hagedornradiusexpression} with \eqref{lightconeradius}, we find that
\bea
\frac{r_H}{r_1} = \frac{r_0}{r_h} \sinh \left(\frac{\beta_H}{2 L}\right)\, e^{\pi r_h/L} \gg 1\,, \label{rhvsr1comp}
\eea
so the infaller hits the Hagedorn wall before encountering even the first image singularity. Therefore, no matter what the precise finite $c$ resolution of the image singularities is, they lie in a region where the semiclassical geometry has already broken down for other reasons (see Figure \ref{infalltimefig}). For a small black hole, no such strict ordering of radii can be established as the location of image singularities can change significantly depending on the values of $r_h/L$.

\section{Discussion: the Fate of an Infalling Observer}

\label{discussionsec}

We have identified several effects that can correct the semiclassical divergences of the infalling correlator. It is natural to ask what these corrections imply for an infalling observer. Let us go through them one by one. The discreteness of the black hole spectrum is not accessible to the infalling observer, and its effects are therefore negligible. The loop corrections coming from the tower of new particles become important only when the infaller has crossed the location of the Hagedorn wall, that is, when $r(\tau)<r_H$. From \eqref{fullhagedornradius}, we can see that 
\bea
\frac{r_H}{r_h} = \sinh \left(\frac{\beta_H}{2 L}\right) \text{csch}\left(\frac{\pi r_h}{L}\right)\,.
\eea
For a large black hole, this number is very small, and these effects remain exponentially suppressed near the horizon. For a small black hole, however, the HKS bound \eqref{HKSbound} allows the Hagedorn wall to lie even outside the horizon. An infalling observer can therefore encounter substantial deviations from the smooth horizon picture when falling into a small black hole, depending on the exact value of $\beta_H/L$ and $r_h/L$.

From \eqref{rhvsr1comp}, it also follows that the large-frequency corrections that resolve the image singularities remain negligible near the horizon of a large black hole. We can therefore conclude that, even after incorporating the corrections that ``resolve'' the semiclassical divergences at the singularity, an infalling observer still sees a smooth horizon in a large BTZ black hole. Thus, an approximate notion of the horizon and (a part of) the black hole interior survives at finite $c$, as one would expect from black hole complementarity \cite{Susskind:1993if}.

It is quite possible that there are other corrections that we have not identified, which become important before the infaller reaches the Hagedorn wall. One possibility is the Horowitz--Polchinski Milne instability \cite{Horowitz:2002mw}. In our calculation, the loop connects a point to its own image. Horowitz and Polchinski considered the gravitational interaction between a particle and its image at a \emph{different} point along the trajectory. This can be much more violent, because the quotient $\gamma$ acts as a
boost in the $X_0$--$X_1$ plane of the embedding space \eqref{boosteq}. Therefore, the probe particle and its $n$-th image are relatively boosted at every
point of the fall, with the
relative rapidity
\bea
   \eta_n \sim \frac{2n\pi r_h}{L}\,.
\eea
Now let us study the gravitational interaction between the particle and its $n$-th image. Following \cite{Horowitz:2002mw}, we model this interaction as the exchange of a shockwave. Suppose the shockwave is emitted at $x(\tau_e)$ and received at a later point $x(\tau_r)$. In the covering space, this corresponds to a shockwave exchanged between $x(\tau_e)$ and $\gamma^n x(\tau_r)$. The center-of-mass energy of this interaction grows exponentially with $n$,
\bea
E_{\text{COM}}\sim \frac{\Delta}{L}\, e^{\eta_n/2} \sim \frac{\Delta}{L}\, e^{\pi n r_h/L}\,.
\eea
For sufficiently large $n$, this exceeds the BTZ threshold,
$G_N E_{\rm COM}\gtrsim 1$, and the backreaction is no longer small.

For our infalling probe, causality places a simple bound on where this can happen. The shock arrives when it has wound $n$ times around the $\phi$ circle:
\bea
\xi\big(x(\tau_e),\gamma^n x(\tau_r)\big)=1\,.
\eea
Now let us make the observation that a later emission results in a later reception of the shockwave. The earliest emission corresponds to a shockwave emitted from $x(0)$. This places $x(\tau_r)$ precisely at the $n$-th image singularity. Therefore, the Horowitz--Polchinski instability associated with the $n$-th image cannot set in before the first image singularity, giving us the bound:
\bea
r_{\rm HP} \leq r_1 \, .
\eea
By \eqref{rhvsr1comp}, $r_1\ll r_H$ for a large black hole. The infaller therefore reaches the Hagedorn wall before its own boosted gravitational
field can return to it, regardless of how violent that return is.

It is natural to ask what the Hagedorn wall corresponds to in a string-theoretic
realisation of the tower. In flat space, modular invariance of the torus amplitude relates the string Hagedorn transition to the winding mode around the thermal circle becoming massless (see, for example, \cite{Atick:1988si}). Therefore, it would be very interesting to check if our interior Hagedorn transition is signalling a tachyonic condensation in the string theory completion. Making this precise is not straightforward. The identification relevant here is along the spatial $\phi$ circle rather than a thermal circle, so the usual argument \cite{Atick:1988si} does not apply directly, and the
existence of a suitable winding mode is delicate \cite{Nekrasov:2002kf,Pioline:2003bs,Rangamani:2007fz}. Whether an actual condensate forms and
replaces the region $r<r_H$, along the lines of \cite{McGreevy:2005ci,Horowitz:2006mr}, is an open question.

We close with a curious observation.\footnote{We thank Chethan Krishnan for pointing out this observation.} For a Euclidean black hole, the thermal circle shrinks to zero size at the tip of the cigar, and the resulting tachyonic condensation implies a Hagedorn-like phase at the horizon in Lorentzian signature (see, for example, \cite{Dabholkar:2001if}). This phase accounts for the black hole entropy and thus provides an explicit realisation of the stretched horizon in the complementary description seen by an asymptotic observer \cite{Susskind:1993if}. The two Hagedorn transitions therefore resolve the semiclassical pathologies seen by the two observers: the smooth horizon for the asymptotic observer and the singularity for the infalling observer. Remarkably, the same tower of particles can underlie both transitions. In the boundary theory, a single number, the gap to the corresponding tower of primaries, then controls both transitions, which suggests that such a tower may be a natural mechanism for implementing quantum aspects of black hole complementarity in holographic black holes. It would be interesting to develop this idea of \emph{Hagedorn complementarity} further.

\acknowledgments
I am grateful to Pawel Caputa, Chethan Krishnan, Matthew M. Roberts, Bo Sundborg, Watse Sybesma, and Lárus Thorlacius for valuable discussions. This work is supported by the ERC Consolidator grant (number: 101125449/acronym: QComplexity). Views and opinions expressed are however those of the author only and do not necessarily reflect those of the European Union or the European Research Council.
Neither the European Union nor the granting authority can be held responsible for them.

%VM is supported by a doctoral grant from the University of Iceland Science Park. 
%This is the most common positions for acknowledgments. A macro is
%available to maintain the same layout and spelling of the heading.

%\paragraph{Note added.} This is also a good position for notes added
%after the paper has been written.

\appendix
\section{Causal structure of the one-loop correction}
\label{ininapp}
Consider the first-order loop correction to the Wightman function $\langle\Phi(x)\Phi(x')\rangle$ in a fixed state coming from the interaction 
\bea
H_{\rm int} = g\int \dd^2 y\, \sqrt{h}\, \Phi^2\Psi^2\,.
\eea
Since we are computing an expectation value rather than an \emph{in-out} scattering amplitude, we need to use the \emph{in-in} formalism \cite{Schwinger:1960qe,Maldacena:2002vr,Weinberg:2005vy}. We will denote the time coordinates at the two points by $t$ and $t'$. The field operator in the Heisenberg picture is given by 
\bea
\Phi_{H}(x) = U^\dagger(t,t_0)\,\Phi_I(x)\,U(t,t_0)\,, \qquad \text{with} \ U = T\exp\big(-i\int_{t_0}^t H_{\rm int}\big)\,.
\eea 
The lower limit of the integral $t_0$ is usually taken to be $-\infty$. We will come back to this point in a bit. Expanding to first order, we get
\bea
\Phi_H(x) = \Phi_I(x) + i\int_{t_0}^{t}\dd t_1\, \big[H_{\rm int}(t_1),\Phi_I(x)\big] + O(g^2)\,.
\eea
The commutator is supported only at $t_1<t$. Therefore, the Heisenberg operator at time $t$ is sensitive to the interaction only in its past. Since $\Psi$ commutes with $\Phi$,
\bea
[\Phi^2(y)\Psi^2(y),\Phi(x)] = 2\Phi(y)\Psi^2(y)[\Phi(y),\Phi(x)]\,.
\eea
As $y$ is forced to be in the past of $x$, the commutator is given by the retarded propagator $[\Phi_I(y),\Phi_I(x)] = i\,G_R(x,y)$. This gives us
\bea
\Phi_H(x) = \Phi_I(x) - \int \dd^3y\,\sqrt{-g}\; G_R(x,y)\, 2g\,\Psi_I^2(y)\Phi_I(y) + O(g^2)\,.
\eea
Now, let us make an important observation. When computing $\langle\Phi_H(x)\Phi_H(x')\rangle$, the integral receives contributions only from points in the past light cones of $x$ and $x'$. Therefore, in our infalling correlator calculation \eqref{bulkloopdiagram}, the intermediate point $y$ must lie in the past light cone of $x(\tau)$. The dominant contribution to the integral, therefore, comes from the configuration where the loop is located at the future insertion point.

One might worry that the loop can extend all the way to the white hole singularity, since these points also lie in the past light cone of the insertion points. In the eternal BTZ geometry, the state is prepared by the Euclidean path integral over half of Euclidean BTZ, and the in--in contour consists of that Euclidean half, followed by Lorentzian evolution forward from the $t=0$ slice and back, followed by the other Euclidean half \cite{Skenderis:2008dg}. The Lorentzian part of the contour, therefore, starts on the Einstein--Rosen bridge and contains no white-hole region.

% Bibliography

\bibliographystyle{JHEP}
\bibliography{refs.bib}

@article{Leutheusser:2021frk,
    author = "Leutheusser, Samuel Aaron Wehlau and Liu, Hong",
    title = "{Emergent Times in Holographic Duality}",
    eprint = "2112.12156",
    archivePrefix = "arXiv",
    primaryClass = "hep-th",
    reportNumber = "MIT-CTP/5382",
    doi = "10.1103/PhysRevD.108.086020",
    journal = "Phys. Rev. D",
    volume = "108",
    number = "8",
    pages = "086020",
    year = "2023"
}

@article{Papadodimas:2012aq,
    author = "Papadodimas, Kyriakos and Raju, Suvrat",
    title = "{An Infalling Observer in AdS/CFT}",
    eprint = "1211.6767",
    archivePrefix = "arXiv",
    primaryClass = "hep-th",
    reportNumber = "HRI-ST-1107, ICTS-2012-10",
    doi = "10.1007/JHEP10(2013)212",
    journal = "JHEP",
    volume = "10",
    pages = "212",
    year = "2013"
}

@article{Maldacena:2001kr,
    author = "Maldacena, Juan Martin",
    title = "{Eternal black holes in anti-de Sitter}",
    eprint = "hep-th/0106112",
    archivePrefix = "arXiv",
    reportNumber = "NSF-ITP-01-59",
    doi = "10.1088/1126-6708/2003/04/021",
    journal = "JHEP",
    volume = "04",
    pages = "021",
    year = "2003"
}

@article{Michel:2019vnk,
    author = "Michel, Ben",
    title = "{Universality in the OPE Coefficients of Holographic 2d CFTs}",
    eprint = "1908.02873",
    archivePrefix = "arXiv",
    primaryClass = "hep-th",
    month = "8",
    year = "2019"
}

@article{Hartman:2014oaa,
    author = "Hartman, Thomas and Keller, Christoph A. and Stoica, Bogdan",
    title = "{Universal Spectrum of 2d Conformal Field Theory in the Large c Limit}",
    eprint = "1405.5137",
    archivePrefix = "arXiv",
    primaryClass = "hep-th",
    reportNumber = "CALT-68-2889, RUNHETC-2014-07",
    doi = "10.1007/JHEP09(2014)118",
    journal = "JHEP",
    volume = "09",
    pages = "118",
    year = "2014"
}

@article{Atick:1988si,
    author = "Atick, Joseph J. and Witten, Edward",
    title = "{The Hagedorn Transition and the Number of Degrees of Freedom of String Theory}",
    reportNumber = "IASSNS-HEP-88-14",
    doi = "10.1016/0550-3213(88)90151-4",
    journal = "Nucl. Phys. B",
    volume = "310",
    pages = "291--334",
    year = "1988"
}

@article{McGreevy:2005ci,
    author = "McGreevy, John and Silverstein, Eva",
    title = "{The Tachyon at the end of the universe}",
    eprint = "hep-th/0506130",
    archivePrefix = "arXiv",
    reportNumber = "SU-ITP-05-22, SLAC-PUB-11283",
    doi = "10.1088/1126-6708/2005/08/090",
    journal = "JHEP",
    volume = "08",
    pages = "090",
    year = "2005"
}

@article{Hijano:2015qja,
    author = "Hijano, Eliot and Kraus, Per and Perlmutter, Eric and Snively, River",
    title = "{Semiclassical Virasoro blocks from AdS$_{3}$ gravity}",
    eprint = "1508.04987",
    archivePrefix = "arXiv",
    primaryClass = "hep-th",
    doi = "10.1007/JHEP12(2015)077",
    journal = "JHEP",
    volume = "12",
    pages = "077",
    year = "2015"
}

@article{Grabovsky:2024jwf,
    author = "Grabovsky, David",
    title = "{Heavy states in 3d gravity and 2d CFT}",
    eprint = "2403.13757",
    archivePrefix = "arXiv",
    primaryClass = "hep-th",
    doi = "10.1007/JHEP07(2024)287",
    journal = "JHEP",
    volume = "07",
    pages = "287",
    year = "2024"
}

@article{Hamilton:2007wj,
    author = "Hamilton, Alex and Kabat, Daniel N. and Lifschytz, Gilad and Lowe, David A.",
    editor = "Sharpe, Eric and Greenspoon, Arthur",
    title = "{Local bulk operators in AdS/CFT and the fate of the BTZ singularity}",
    eprint = "0710.4334",
    archivePrefix = "arXiv",
    primaryClass = "hep-th",
    journal = "AMS/IP Stud. Adv. Math.",
    volume = "44",
    pages = "85--100",
    year = "2008"
}

@article{Fidkowski:2003nf,
    author = "Fidkowski, Lukasz and Hubeny, Veronika and Kleban, Matthew and Shenker, Stephen",
    title = "{The Black hole singularity in AdS / CFT}",
    eprint = "hep-th/0306170",
    archivePrefix = "arXiv",
    reportNumber = "SU-ITP-03-16",
    doi = "10.1088/1126-6708/2004/02/014",
    journal = "JHEP",
    volume = "02",
    pages = "014",
    year = "2004"
}

@article{Faulkner:2017hll,
    author = "Faulkner, Thomas and Wang, Huajia",
    title = "{Probing beyond ETH at large $c$}",
    eprint = "1712.03464",
    archivePrefix = "arXiv",
    primaryClass = "hep-th",
    doi = "10.1007/JHEP06(2018)123",
    journal = "JHEP",
    volume = "06",
    pages = "123",
    year = "2018"
}

@article{Fitzpatrick:2016ive,
    author = "Fitzpatrick, A. Liam and Kaplan, Jared and Li, Daliang and Wang, Junpu",
    title = "{On information loss in AdS$_{3}$/CFT$_{2}$}",
    eprint = "1603.08925",
    archivePrefix = "arXiv",
    primaryClass = "hep-th",
    doi = "10.1007/JHEP05(2016)109",
    journal = "JHEP",
    volume = "05",
    pages = "109",
    year = "2016"
}

@article{Carlip:1995qv,
    author = "Carlip, Steven",
    title = "{The (2+1)-Dimensional black hole}",
    eprint = "gr-qc/9506079",
    archivePrefix = "arXiv",
    reportNumber = "UCD-95-15",
    doi = "10.1088/0264-9381/12/12/005",
    journal = "Class. Quant. Grav.",
    volume = "12",
    pages = "2853--2880",
    year = "1995"
}

@article{Sarkar:2020yjs,
    author = "Sarkar, Debajyoti and Visser, Manus",
    title = "{The first law of differential entropy and holographic complexity}",
    eprint = "2008.12673",
    archivePrefix = "arXiv",
    primaryClass = "hep-th",
    doi = "10.1007/JHEP11(2020)004",
    journal = "JHEP",
    volume = "11",
    pages = "004",
    year = "2020"
}

@article{Sathiapalan:1986db,
    author = "Sathiapalan, B.",
    title = "{Vortices on the String World Sheet and Constraints on Toral Compactification}",
    reportNumber = "UCLA/86/TEP/37",
    doi = "10.1103/PhysRevD.35.3277",
    journal = "Phys. Rev. D",
    volume = "35",
    pages = "3277",
    year = "1987"
}

@article{Kogan:1987jd,
    author = "Kogan, Ya. I.",
    editor = "Shifman, M. and Vainshtein, A. and Wheater, J.",
    title = "{Vortices on the World Sheet and String's Critical Dynamics}",
    reportNumber = "ITEP-87-110",
    journal = "JETP Lett.",
    volume = "45",
    pages = "709--712",
    year = "1987"
}

@article{OBrien:1987kzw,
    author = "O'Brien, K. H. and Tan, C. I.",
    title = "{Modular Invariance of Thermopartition Function and Global Phase Structure of Heterotic String}",
    reportNumber = "BROWN-HET-602",
    doi = "10.1103/PhysRevD.36.1184",
    journal = "Phys. Rev. D",
    volume = "36",
    pages = "1184",
    year = "1987"
}

@article{Horowitz:2002mw,
    author = "Horowitz, Gary T. and Polchinski, Joseph",
    title = "{Instability of space - like and null orbifold singularities}",
    eprint = "hep-th/0206228",
    archivePrefix = "arXiv",
    doi = "10.1103/PhysRevD.66.103512",
    journal = "Phys. Rev. D",
    volume = "66",
    pages = "103512",
    year = "2002"
}

@article{Nekrasov:2002kf,
    author = "Nekrasov, Nikita A.",
    editor = "Kaidalov, A. B. and Vysotsky, M. I.",
    title = "{Milne universe, tachyons, and quantum group}",
    eprint = "hep-th/0203112",
    archivePrefix = "arXiv",
    reportNumber = "IHES-P-02-13, ITEP-TH-14-02",
    doi = "10.1080/0142241021000054176",
    journal = "Surveys High Energ. Phys.",
    volume = "17",
    pages = "115--124",
    year = "2002"
}

@article{Pioline:2003bs,
    author = "Pioline, B. and Berkooz, M.",
    title = "{Strings in an electric field, and the Milne universe}",
    eprint = "hep-th/0307280",
    archivePrefix = "arXiv",
    reportNumber = "LPTHE-03-21, WIS-20-03-DPP",
    doi = "10.1088/1475-7516/2003/11/007",
    journal = "JCAP",
    volume = "11",
    pages = "007",
    year = "2003"
}

@article{Rangamani:2007fz,
    author = "Rangamani, Mukund and Ross, Simon F.",
    title = "{Winding tachyons in BTZ}",
    eprint = "0706.0663",
    archivePrefix = "arXiv",
    primaryClass = "hep-th",
    reportNumber = "DCPT-07-21",
    doi = "10.1103/PhysRevD.77.026010",
    journal = "Phys. Rev. D",
    volume = "77",
    pages = "026010",
    year = "2008"
}

@article{Kraus:2002iv,
    author = "Kraus, Per and Ooguri, Hirosi and Shenker, Stephen",
    title = "{Inside the horizon with AdS / CFT}",
    eprint = "hep-th/0212277",
    archivePrefix = "arXiv",
    reportNumber = "UCLA-02-TEP-41, CALT-68-2421, SU-ITP-02-45",
    doi = "10.1103/PhysRevD.67.124022",
    journal = "Phys. Rev. D",
    volume = "67",
    pages = "124022",
    year = "2003"
}

@article{Susskind:1993if,
    author = "Susskind, Leonard and Thorlacius, Larus and Uglum, John",
    title = "{The Stretched horizon and black hole complementarity}",
    eprint = "hep-th/9306069",
    archivePrefix = "arXiv",
    reportNumber = "SU-ITP-93-15",
    doi = "10.1103/PhysRevD.48.3743",
    journal = "Phys. Rev. D",
    volume = "48",
    pages = "3743--3761",
    year = "1993"
}

@article{Almheiri:2012rt,
    author = "Almheiri, Ahmed and Marolf, Donald and Polchinski, Joseph and Sully, James",
    title = "{Black Holes: Complementarity or Firewalls?}",
    eprint = "1207.3123",
    archivePrefix = "arXiv",
    primaryClass = "hep-th",
    doi = "10.1007/JHEP02(2013)062",
    journal = "JHEP",
    volume = "02",
    pages = "062",
    year = "2013"
}

@article{Marolf:2013dba,
    author = "Marolf, Donald and Polchinski, Joseph",
    title = "{Gauge/Gravity Duality and the Black Hole Interior}",
    eprint = "1307.4706",
    archivePrefix = "arXiv",
    primaryClass = "hep-th",
    doi = "10.1103/PhysRevLett.111.171301",
    journal = "Phys. Rev. Lett.",
    volume = "111",
    pages = "171301",
    year = "2013"
}

@article{Papadodimas:2013wnh,
    author = "Papadodimas, Kyriakos and Raju, Suvrat",
    title = "{Black Hole Interior in the Holographic Correspondence and the Information Paradox}",
    eprint = "1310.6334",
    archivePrefix = "arXiv",
    primaryClass = "hep-th",
    reportNumber = "ICTS-2013-20, ICTS/2013/20",
    doi = "10.1103/PhysRevLett.112.051301",
    journal = "Phys. Rev. Lett.",
    volume = "112",
    number = "5",
    pages = "051301",
    year = "2014"
}

@article{Papadodimas:2013jku,
    author = "Papadodimas, Kyriakos and Raju, Suvrat",
    title = "{State-Dependent Bulk-Boundary Maps and Black Hole Complementarity}",
    eprint = "1310.6335",
    archivePrefix = "arXiv",
    primaryClass = "hep-th",
    reportNumber = "ICTS-2013-21, ICTS/2013/21",
    doi = "10.1103/PhysRevD.89.086010",
    journal = "Phys. Rev. D",
    volume = "89",
    number = "8",
    pages = "086010",
    year = "2014"
}

@article{Hamilton:2006fh,
    author = "Hamilton, Alex and Kabat, Daniel N. and Lifschytz, Gilad and Lowe, David A.",
    title = "{Local bulk operators in AdS/CFT: A Holographic description of the black hole interior}",
    eprint = "hep-th/0612053",
    archivePrefix = "arXiv",
    reportNumber = "CU-TP-1162",
    doi = "10.1103/PhysRevD.75.106001",
    journal = "Phys. Rev. D",
    volume = "75",
    pages = "106001",
    year = "2007",
    note = "[Erratum: Phys.Rev.D 75, 129902 (2007)]"
}

@article{Almheiri:2017fbd,
    author = "Almheiri, Ahmed and Anous, Tarek and Lewkowycz, Aitor",
    title = "{Inside out: meet the operators inside the horizon. On bulk reconstruction behind causal horizons}",
    eprint = "1707.06622",
    archivePrefix = "arXiv",
    primaryClass = "hep-th",
    doi = "10.1007/JHEP01(2018)028",
    journal = "JHEP",
    volume = "01",
    pages = "028",
    year = "2018"
}

@article{Jafferis:2020ora,
    author = "Jafferis, Daniel Louis and Lamprou, Lampros",
    title = "{Inside the hologram: reconstructing the bulk observer{\textquoteright}s experience}",
    eprint = "2009.04476",
    archivePrefix = "arXiv",
    primaryClass = "hep-th",
    doi = "10.1007/JHEP03(2022)084",
    journal = "JHEP",
    volume = "03",
    pages = "084",
    year = "2022"
}

@article{Gao:2021tzr,
    author = "Gao, Ping and Lamprou, Lampros",
    title = "{Seeing behind black hole horizons in SYK}",
    eprint = "2111.14010",
    archivePrefix = "arXiv",
    primaryClass = "hep-th",
    doi = "10.1007/JHEP06(2022)143",
    journal = "JHEP",
    volume = "06",
    pages = "143",
    year = "2022"
}

@article{deBoer:2022zps,
    author = "de Boer, Jan and Jafferis, Daniel Louis and Lamprou, Lampros",
    title = "{On black hole interior reconstruction, singularities and the emergence of time}",
    eprint = "2211.16512",
    archivePrefix = "arXiv",
    primaryClass = "hep-th",
    month = "11",
    year = "2022"
}

@article{Lowe:2014vfa,
    author = "Lowe, David A. and Thorlacius, Larus",
    title = "{Black hole complementarity: The inside view}",
    eprint = "1402.4545",
    archivePrefix = "arXiv",
    primaryClass = "hep-th",
    reportNumber = "BROWN-HET-1655, NORDITA-2014-18",
    doi = "10.1016/j.physletb.2014.08.062",
    journal = "Phys. Lett. B",
    volume = "737",
    pages = "320--324",
    year = "2014"
}

@article{Larjo:2012jt,
    author = "Larjo, Klaus and Lowe, David A. and Thorlacius, Larus",
    title = "{Black holes without firewalls}",
    eprint = "1211.4620",
    archivePrefix = "arXiv",
    primaryClass = "hep-th",
    reportNumber = "BROWN-HET-1636, NORDITA-2012-90, RH-10-2012",
    doi = "10.1103/PhysRevD.87.104018",
    journal = "Phys. Rev. D",
    volume = "87",
    number = "10",
    pages = "104018",
    year = "2013"
}

@article{Franken:2026lgp,
    author = "Franken, Victor and Mertens, Thomas G. and de S. L. Torres, Bruno",
    title = "{The endless journey towards the horizon of a quantum black hole}",
    eprint = "2609.05164",
    archivePrefix = "arXiv",
    primaryClass = "hep-th",
    month = "9",
    year = "2026"
}

@article{Banados:1992wn,
    author = "Banados, Maximo and Teitelboim, Claudio and Zanelli, Jorge",
    title = "{The Black hole in three-dimensional space-time}",
    eprint = "hep-th/9204099",
    archivePrefix = "arXiv",
    reportNumber = "PRINT-92-0151 (CHILE), IASSNS-HEP-92-29",
    doi = "10.1103/PhysRevLett.69.1849",
    journal = "Phys. Rev. Lett.",
    volume = "69",
    pages = "1849--1851",
    year = "1992"
}

@article{Banados:1992gq,
    author = "Banados, Maximo and Henneaux, Marc and Teitelboim, Claudio and Zanelli, Jorge",
    title = "{Geometry of the (2+1) black hole}",
    eprint = "gr-qc/9302012",
    archivePrefix = "arXiv",
    reportNumber = "IASSNS-HEP-92-81",
    doi = "10.1103/PhysRevD.48.1506",
    journal = "Phys. Rev. D",
    volume = "48",
    pages = "1506--1525",
    year = "1993",
    note = "[Erratum: Phys.Rev.D 88, 069902 (2013)]"
}

@article{Hamilton:2005ju,
    author = "Hamilton, Alex and Kabat, Daniel N. and Lifschytz, Gilad and Lowe, David A.",
    title = "{Local bulk operators in AdS/CFT: A Boundary view of horizons and locality}",
    eprint = "hep-th/0506118",
    archivePrefix = "arXiv",
    reportNumber = "BROWN-HET-1448, CU-TP-1130",
    doi = "10.1103/PhysRevD.73.086003",
    journal = "Phys. Rev. D",
    volume = "73",
    pages = "086003",
    year = "2006"
}

@article{Fitzpatrick:2015zha,
    author = "Fitzpatrick, A. Liam and Kaplan, Jared and Walters, Matthew T.",
    title = "{Virasoro Conformal Blocks and Thermality from Classical Background Fields}",
    eprint = "1501.05315",
    archivePrefix = "arXiv",
    primaryClass = "hep-th",
    doi = "10.1007/JHEP11(2015)200",
    journal = "JHEP",
    volume = "11",
    pages = "200",
    year = "2015"
}

@article{El-Showk:2011yvt,
    author = "El-Showk, Sheer and Papadodimas, Kyriakos",
    title = "{Emergent Spacetime and Holographic CFTs}",
    eprint = "1101.4163",
    archivePrefix = "arXiv",
    primaryClass = "hep-th",
    doi = "10.1007/JHEP10(2012)106",
    journal = "JHEP",
    volume = "10",
    pages = "106",
    year = "2012"
}

@article{Heemskerk:2009pn,
    author = "Heemskerk, Idse and Penedones, Joao and Polchinski, Joseph and Sully, James",
    title = "{Holography from Conformal Field Theory}",
    eprint = "0907.0151",
    archivePrefix = "arXiv",
    primaryClass = "hep-th",
    reportNumber = "NSF-KITP-09-110",
    doi = "10.1088/1126-6708/2009/10/079",
    journal = "JHEP",
    volume = "10",
    pages = "079",
    year = "2009"
}

@article{Belin:2014fna,
    author = "Belin, Alexandre and Keller, Christoph A. and Maloney, Alexander",
    title = "{String Universality for Permutation Orbifolds}",
    eprint = "1412.7159",
    archivePrefix = "arXiv",
    primaryClass = "hep-th",
    doi = "10.1103/PhysRevD.91.106005",
    journal = "Phys. Rev. D",
    volume = "91",
    number = "10",
    pages = "106005",
    year = "2015"
}

@article{Festuccia:2005pi,
    author = "Festuccia, Guido and Liu, Hong",
    title = "{Excursions beyond the horizon: Black hole singularities in Yang-Mills theories. I.}",
    eprint = "hep-th/0506202",
    archivePrefix = "arXiv",
    reportNumber = "MIT-CTP-3641",
    doi = "10.1088/1126-6708/2006/04/044",
    journal = "JHEP",
    volume = "04",
    pages = "044",
    year = "2006"
}

@article{Grinberg:2020fdj,
    author = "Grinberg, Matan and Maldacena, Juan",
    title = "{Proper time to the black hole singularity from thermal one-point functions}",
    eprint = "2011.01004",
    archivePrefix = "arXiv",
    primaryClass = "hep-th",
    doi = "10.1007/JHEP03(2021)131",
    journal = "JHEP",
    volume = "03",
    pages = "131",
    year = "2021"
}

@article{Horowitz:2023ury,
    author = "Horowitz, Gary T. and Leung, Henry and Queimada, Leonel and Zhao, Ying",
    title = "{Boundary signature of singularity in the presence of a shock wave}",
    eprint = "2310.03076",
    archivePrefix = "arXiv",
    primaryClass = "hep-th",
    doi = "10.21468/SciPostPhys.16.2.060",
    journal = "SciPost Phys.",
    volume = "16",
    number = "2",
    pages = "060",
    year = "2024"
}

@article{Ceplak:2024bja,
    author = "{\v{C}}eplak, Nejc and Liu, Hong and Parnachev, Andrei and Valach, Samuel",
    title = "{Black hole singularity from OPE}",
    eprint = "2404.17286",
    archivePrefix = "arXiv",
    primaryClass = "hep-th",
    doi = "10.1007/JHEP10(2024)105",
    journal = "JHEP",
    volume = "10",
    pages = "105",
    year = "2024"
}

@article{Hubeny:2006yu,
    author = "Hubeny, Veronika E and Liu, Hong and Rangamani, Mukund",
    title = "{Bulk-cone singularities {\&} signatures of horizon formation in AdS/CFT}",
    eprint = "hep-th/0610041",
    archivePrefix = "arXiv",
    reportNumber = "DCPT-06-29, MIT-CTP-3775",
    doi = "10.1088/1126-6708/2007/01/009",
    journal = "JHEP",
    volume = "01",
    pages = "009",
    year = "2007"
}

@article{Dodelson:2020lal,
    author = "Dodelson, Matthew and Ooguri, Hirosi",
    title = "{Singularities of thermal correlators at strong coupling}",
    eprint = "2010.09734",
    archivePrefix = "arXiv",
    primaryClass = "hep-th",
    doi = "10.1103/PhysRevD.103.066018",
    journal = "Phys. Rev. D",
    volume = "103",
    number = "6",
    pages = "066018",
    year = "2021"
}

@article{Hamilton:2006az,
    author = "Hamilton, Alex and Kabat, Daniel N. and Lifschytz, Gilad and Lowe, David A.",
    title = "{Holographic representation of local bulk operators}",
    eprint = "hep-th/0606141",
    archivePrefix = "arXiv",
    reportNumber = "CU-TP-1149",
    doi = "10.1103/PhysRevD.74.066009",
    journal = "Phys. Rev. D",
    volume = "74",
    pages = "066009",
    year = "2006"
}

@article{Kraus:2016nwo,
    author = "Kraus, Per and Maloney, Alexander",
    title = "{A cardy formula for three-point coefficients or how the black hole got its spots}",
    eprint = "1608.03284",
    archivePrefix = "arXiv",
    primaryClass = "hep-th",
    doi = "10.1007/JHEP05(2017)160",
    journal = "JHEP",
    volume = "05",
    pages = "160",
    year = "2017"
}

@article{Horowitz:2006mr,
    author = "Horowitz, Gary T. and Silverstein, Eva",
    title = "{The Inside story: Quasilocal tachyons and black holes}",
    eprint = "hep-th/0601032",
    archivePrefix = "arXiv",
    reportNumber = "SLAC-PUB-11616, SU-ITP-06-01",
    doi = "10.1103/PhysRevD.73.064016",
    journal = "Phys. Rev. D",
    volume = "73",
    pages = "064016",
    year = "2006"
}

@article{Frenkel:2020ysx,
    author = "Frenkel, Alexander and Hartnoll, Sean A. and Kruthoff, Jorrit and Shi, Zhengyan D.",
    title = "{Holographic flows from CFT to the Kasner universe}",
    eprint = "2004.01192",
    archivePrefix = "arXiv",
    primaryClass = "hep-th",
    doi = "10.1007/JHEP08(2020)003",
    journal = "JHEP",
    volume = "08",
    pages = "003",
    year = "2020"
}

@article{Schwinger:1960qe,
    author = "Schwinger, Julian S.",
    title = "{Brownian motion of a quantum oscillator}",
    doi = "10.1063/1.1703727",
    journal = "J. Math. Phys.",
    volume = "2",
    pages = "407--432",
    year = "1961"
}

@article{Maldacena:2002vr,
    author = "Maldacena, Juan Martin",
    title = "{Non-Gaussian features of primordial fluctuations in single field inflationary models}",
    eprint = "astro-ph/0210603",
    archivePrefix = "arXiv",
    doi = "10.1088/1126-6708/2003/05/013",
    journal = "JHEP",
    volume = "05",
    pages = "013",
    year = "2003"
}

@article{Weinberg:2005vy,
    author = "Weinberg, Steven",
    title = "{Quantum contributions to cosmological correlations}",
    eprint = "hep-th/0506236",
    archivePrefix = "arXiv",
    reportNumber = "UTTG-01-05",
    doi = "10.1103/PhysRevD.72.043514",
    journal = "Phys. Rev. D",
    volume = "72",
    pages = "043514",
    year = "2005"
}

@article{Skenderis:2008dg,
    author = "Skenderis, Kostas and van Rees, Balt C.",
    title = "{Real-time gauge/gravity duality: Prescription, Renormalization and Examples}",
    eprint = "0812.2909",
    archivePrefix = "arXiv",
    primaryClass = "hep-th",
    reportNumber = "ITFA-2008-50",
    doi = "10.1088/1126-6708/2009/05/085",
    journal = "JHEP",
    volume = "05",
    pages = "085",
    year = "2009"
}

@article{Dabholkar:2001if,
    author = "Dabholkar, Atish",
    title = "{Tachyon condensation and black hole entropy}",
    eprint = "hep-th/0111004",
    archivePrefix = "arXiv",
    reportNumber = "HUPT-01-A051, TIFR-TH-01-42",
    doi = "10.1103/PhysRevLett.88.091301",
    journal = "Phys. Rev. Lett.",
    volume = "88",
    pages = "091301",
    year = "2002"
}

@article{Goto:2017olq,
    author = "Goto, Kanato and Takayanagi, Tadashi",
    title = "{CFT descriptions of bulk local states in the AdS black holes}",
    eprint = "1704.00053",
    archivePrefix = "arXiv",
    primaryClass = "hep-th",
    reportNumber = "UT-Komaba17-2, YITP-17-31, IPMU17-0047",
    doi = "10.1007/JHEP10(2017)153",
    journal = "JHEP",
    volume = "10",
    pages = "153",
    year = "2017"
}

@article{Dodelson:2023nnr,
    author = "Dodelson, Matthew and Iossa, Cristoforo and Karlsson, Robin and Lupsasca, Alexandru and Zhiboedov, Alexander",
    title = "{Black hole bulk-cone singularities}",
    eprint = "2310.15236",
    archivePrefix = "arXiv",
    primaryClass = "hep-th",
    reportNumber = "CERN-TH-2023-192",
    doi = "10.1007/JHEP07(2024)046",
    journal = "JHEP",
    volume = "07",
    pages = "046",
    year = "2024"
}

@article{Iliesiu:2018fao,
    author = "Iliesiu, Luca and Kolo{\u{g}}lu, Murat and Mahajan, Raghu and Perlmutter, Eric and Simmons-Duffin, David",
    title = "{The Conformal Bootstrap at Finite Temperature}",
    eprint = "1802.10266",
    archivePrefix = "arXiv",
    primaryClass = "hep-th",
    reportNumber = "CALT-TH-2018-013, PUPT-2550",
    doi = "10.1007/JHEP10(2018)070",
    journal = "JHEP",
    volume = "10",
    pages = "070",
    year = "2018"
}

@article{Grozdanov:2026ktq,
    author = "Grozdanov, Sa{\v{s}}o and Movrin, Vita and Valach, Samuel",
    title = "{Bouncing Geodesics, Singularities, and the Cavity Thermal Product Formula in Asymptotically Flat and de Sitter Black Holes}",
    eprint = "2606.11297",
    archivePrefix = "arXiv",
    primaryClass = "hep-th",
    month = "6",
    year = "2026"
}

@article{Coviello:2026tol,
    author = "Coviello, Chiara and Gupta, Ansh and Hennigar, Robie A. and Shi, Kai and Svesko, Andrew",
    title = "{Looking inside a quantum black hole}",
    eprint = "2609.20788",
    archivePrefix = "arXiv",
    primaryClass = "hep-th",
    month = "9",
    year = "2026"
}

@article{Coviello:2026hdc,
    author = "Coviello, Chiara and Gupta, Ansh and Hennigar, Robie A. and Shi, Kai and Svesko, Andrew",
    title = "{Quasinormal modes and quantum black hole interiors}",
    eprint = "2609.26724",
    archivePrefix = "arXiv",
    primaryClass = "hep-th",
    month = "9",
    year = "2026"
}

\end{document}